# Second Harmonic Generation in Chemical Vapor Deposition Synthesized CuS Crystals

Abdulsalam Aji Suleiman[a]*, Reza Rahighi[a], Amir Parsi[a], and Talip Serkan Kasirga[a,b]*

[a]Institute of Materials Science and Nanotechnology, Bilkent University UNAM, Ankara 06800, Turkey

[b]Department of Physics, Bilkent University, Ankara 06800, Turkey

*Corresponding authors;

Email: kasirga@unam.bilkent.edu.tr; abdulsalam@unam.bilkent.edu.tr

**ABSTRACT**

Copper sulfide (CuS) has garnered significant attention in various fields of application due to its unique electronic, optical, and catalytic features. In this study, we present the chemical vapor deposition (CVD)-based synthesis of ultrathin CuS crystals as thin as 14 nm with lateral sizes up to 60 μm. The structure, morphology, and composition of the as-synthesized CuS crystals were thoroughly characterized. Among our results, we measured the first-order temperature coefficients of Raman shifts of CuS. Moreover, we showed that CuS crystals exhibited an unexpected second harmonic generation (SHG), which is attributed to the presence of defects in the CuS lattice. Our results suggest that single crystalline CuS possesses a considerable potential for nonlinear optical applications in conjunction with its current applications in electronics and catalysis.

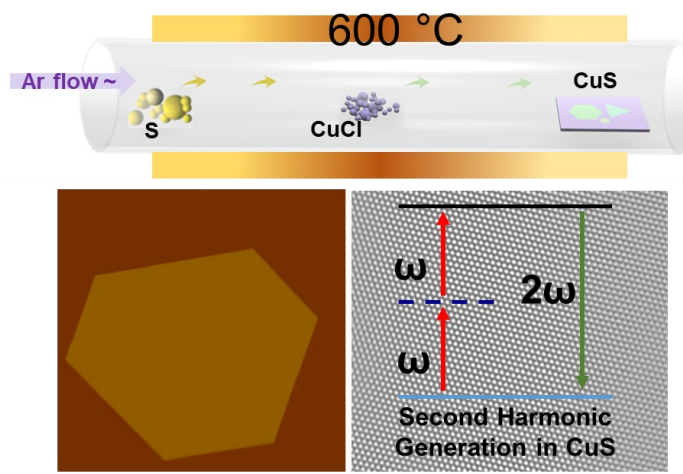

**Introduction**

Copper sulfide (CuS) is an interesting binary compound owing to its exceptional optical, electronic, and catalytic characteristics.[1-6] It is a multifaceted material that manifests in diverse crystal structures, such as hexagonal covellite and orthorhombic anilite, among others.[7-10] The distinctive attributes of CuS crystals have resulted in a diverse range of applications, including but not limited to energy storage sensors, photovoltaic devices, and thermoelectric devices.[11-15] The crystals of CuS exhibit remarkable optical characteristics, such as elevated quantum yields of photoluminescence and exceptional absorption of visible light. The material's robust light absorption properties have been utilized in various photovoltaic devices, including but not limited to dye-sensitized solar cells and thin-film solar cells, to optimize their efficiency.[16, 17] Wet-chemistry techniques have been employed to synthesize CuS nanoparticles, which exhibit considerable potential as lithium/magnesium-ion battery components owing to their superior energy storage capabilities.[18, 19] Furthermore, the photocatalytic characteristics of CuS have been investigated, wherein its considerable specific surface area has been identified as a pivotal factor in augmenting photocatalytic efficacy.[20] The characteristic feature of CuS renders it a viable contender for deployment in the domains of environmental remediation and hydrogen generation via water electrolysis. As reported by As reported by Fan et al. the light-harvesting and charge dissociation functionalities of heterostructured $ZnIn_2S_4$/CuS nanosheets can be considerably enhanced without requiring a co-catalyst.[21] The observed improvement can be attributed to the robust interplay between the assembled p-n heterostructures, which effectively promotes the transfer and separation of charges. Zhu et al. recently showed the outstanding efficiency of CuS in the oxygen evolution reaction, exhibiting minimal overpotential and exceptional electrolytic endurance.[22] Although there are many reports on the energy storage and photocatalysis applications of CuS, its physicochemical properties have not been thoroughly investigated. Consequently, a plethora of unexplored possibilities exist for uncovering novel CuS characteristics and uses in diverse domains, including energy conversion and storage, sensing, and optoelectronics. Further exploration of the characteristics of CuS and its analogous compounds has the potential to facilitate significant progress in the field of materials science and the innovation of advanced technologies.

In this work, CuS crystals were synthesized using a single-step CVD technique at a relatively low temperature of 600 °C. The as-synthesized CuS crystal's structural, morphological, and compositional features were thoroughly characterized. Intriguingly, these CuS crystals exhibited an unexpected second harmonic generation (SHG), which is attributed to the presence of defects or lattice distortions within the CuS crystal structure. Such lattice imperfections led to a breaking of the intrinsic centrosymmetric nature of CuS crystals, which resulted in the SHG response. The as-synthesized CuS crystal, with its unexpected SHG response and defect-related properties, may pave the way for novel nonlinear optical materials that can be employed in various applications across multiple fields, including photonics, telecoms, and sensing technologies.

**Results and Discussion**

Copper(I) chloride (CuCl) powder was opted as the Cu source for growth of CuS crystals, due to its relatively low melting point temperature. A small, asymmetric crucible was chosen to this end, filled with scant amount of CuCl powder, and put in middle of tubular CVD furnace as can be seen in the schematic setup in the supporting information (Figure S1). During the synthesis process, the optimum growth temperature was found to be about 600 °C, lower than the other CVD synthesis of copper-based chalcogenides.[23, 24] Additional information regarding the CVD growth process is provided in the experimental section. **Figure 1a** shows a typical optical micrograph of CuS crystals grown on a mica. The lateral length of the grown ultrathin CuS crystals can be up to 60 μm (**Figure 1b**). Mica is used as a substrate because of its atomic-level smooth and inert surface, which has been widely reported as a favorable substrate for ultrathin material synthesis.[25, 26] We used different substrates for comparison and substrate effect influence, including $SiO_2$ and bare Si, and found that mica worked best in terms of crystal homogeneity, thickness, and size. Figure S2 shows the results of the as-grown CuS crystals on several substrates. Atomic force microscopy (AFM) height trace map given in **Figure 1c** confirms that the surface of CuS is very smooth and the thickness of the studied crystal is found to be about 14 nm. The X-ray diffraction (XRD) θ-2θ scan of the transferred CuS crystals on the $SiO_2$/Si substrate, shown in **Figure 1d**, unambiguously corresponds to the hexagonal phase of CuS (PDF No. 06-0464).[27] The pronounced characteristic peaks corresponding to the (002), (006), and (008) planes indicate that CuS crystals predominantly grow along the basal plane (00*l*). To further investigate the crystallographic orientation of the CuS crystals, electron backscatter diffraction (EBSD) was used. This technique is highly effective in characterizing the microstructure of materials.[28, 29] As seen in **Figure 1e**, the EBSD inverse pole figure (IPF) map exhibits a consistent color contrast within the hexagonal domains along the basal plane of CuS ([00l] direction). This observation implies that the hexagonal CuS crystal possesses a single-crystalline nature and a well-ordered in-plane orientation, which is in agreement with the XRD findings. However, the pole figure (PF) corresponding to the (0001) projection plane (**Figure 1f**) reveals an impaired [0001] out-of-plane orientation. This finding suggests the existence of defects or lattice distortions within the CuS crystal structure, which may influence its properties.[22, 30]

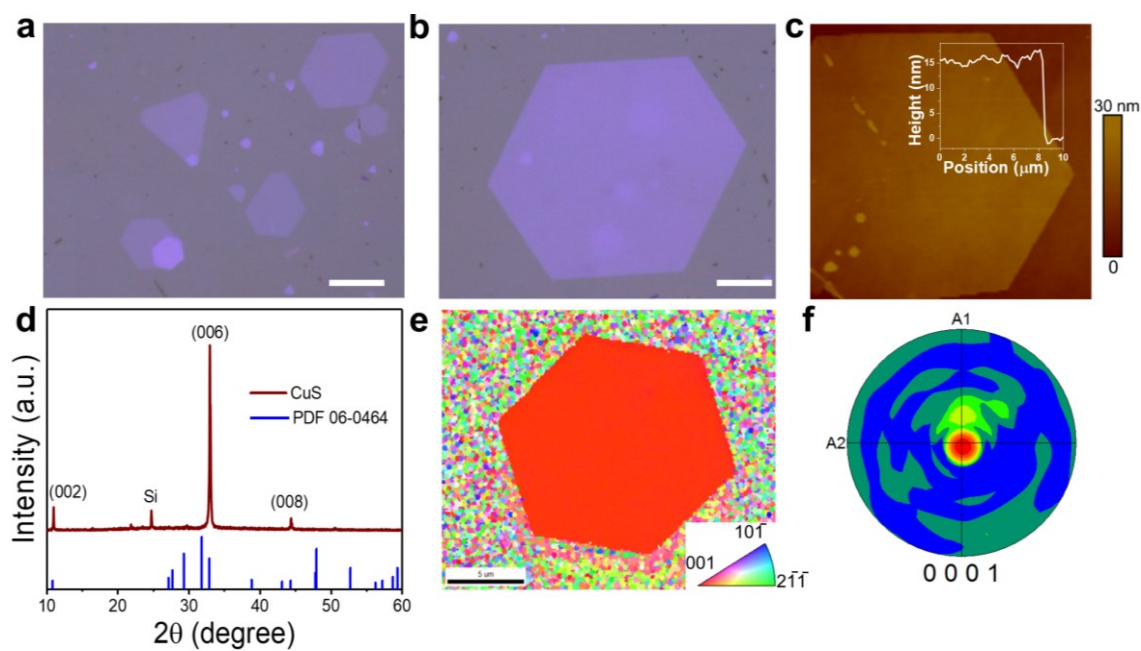

**Figure 1:** Optical, morphological and structural characterization of CuS. **(a-b)** Typical OM image of the as-grown CuS crystals, the scale bars are 10 and 20 μm, respectively. **(c)** AFM image of the CuS crystal and its height profile of the in the inset. **(d)** XRD pattern of CuS crystals on SiO$_2$/Si substrate. **(e)** EBSD inverse pole figure (IPF) map along the c-axis of CuS crystal on SiO$_2$/Si substrate the length of the scale bar corresponds to 5 μm. **(f)** A [0001] pole figure (PF) that corresponds to the IPF map in (e).

Raman spectroscopy with a 532 nm excitation laser was used to investigate the intrinsic properties and identify the fingerprint of the CuS crystal structure. As shown in **Figure 2a**, the Raman spectrum of the CuS crystal shows four distinct Raman peaks at 90.1, 130, 279, and 471

cm$^{-1}$, representing $E_{2g}$, $A_1$, $E_g^1$, and $A_1$ modes, respectively. Among these peaks, the strong characteristic peak at 471.0 cm$^{-1}$ can be attributed to the stretching mode of the S-S bond, corresponding to the $S_2$ groups of the recognized crystal structure of CuS lattice. The spatially resolved Raman mapping images (**Figure 2b**) of the four characteristic peaks (60, 138, 267, and 471 cm$^{-1}$) exhibit uniformity throughout the crystalline sheet of CuS. Temperature-dependent Raman spectroscopy is employed to study the atomic bonding and thermal expansion of the crystals.[31] **Figure 2c** shows the typical temperature-dependent Raman spectra for the grown CuS crystal at temperatures ranging from 80 to 300 K. Temperature-dependent Raman spectroscopy of CuS crystal provides valuable information about its thermal behavior as vibrational modes in its lattice structure. The relationship between temperature and CuS Raman modes can be expressed through a linear equation: $\omega(T) = \omega_0 + \chi T$, where $\omega_0$, T, and $\chi$ are the Raman peak position at 0 K, the Kelvin temperature, and the first-order temperature coefficient, respectively. Raman modes at 60, 138, 267, and 471 cm$^{-1}$ labeled P1, P2, P3, and P4 each exhibit temperature coefficients ($\chi$-values) that fall within the following parameters:-0.00662, -0.02677, 0.025673, and 0.02535 cm$^{-1}$ K$^{-1}$ respectively. Negative $\chi$-values indicate that Raman mode frequencies decrease with temperature increases, suggesting an anharmonic behavior of lattice vibrations. The P1 mode exhibits relatively lower anharmonic behavior than other modes, suggesting temperature variations may affect it less significant as, P2, P3, and P4 modes showed larger negative $\chi$ values suggesting stronger anharmonic behavior (**Figure 2d**). This evidence suggests that these modes are susceptible to temperature variations and that thermal expansion or changes in bonding strength in response to temperature changes may have an impact.[32, 33] Also, previous studies have shown that the dominant temperature coefficient of low-dimensional materials is linked to the van der Waals interaction between layers, and this is often used to explain why the Raman peak shifting changes with temperature.[34-36] The $\chi$-values mentioned above offer valuable information regarding the thermal characteristics of CuS, which could facilitate comprehension of its properties and potential utilization in devices that are sensitive to temperature.

.

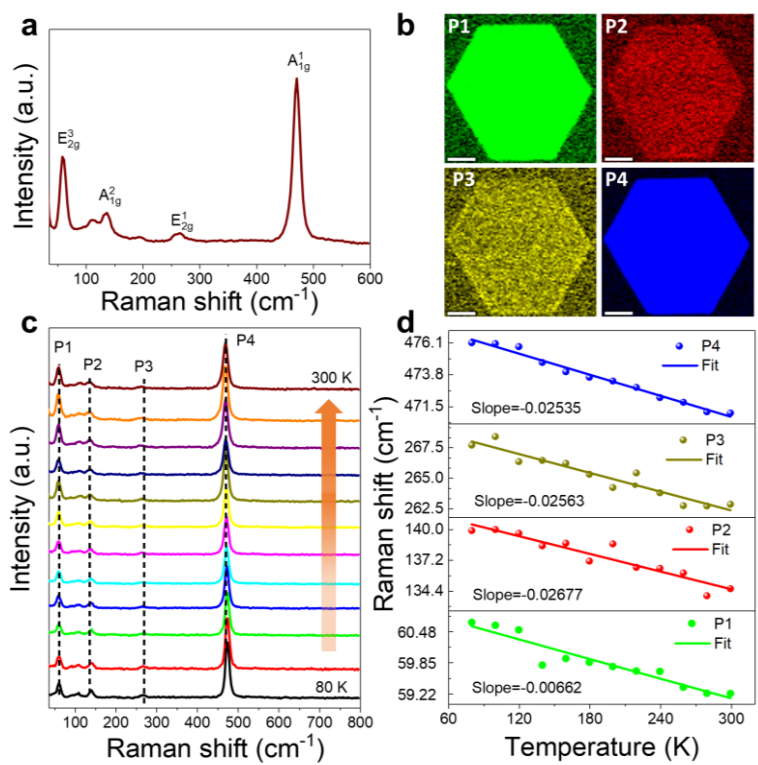

**Figure 2:** Raman characterization of CuS. **(a)** Raman spectrum of the CuS lattice on mica substrate. **(b)** Spatially resolved Raman mapping images of the CuS characteristic peaks $E_{2g}^3$ (P1), $A_{1g}^2$ (P2), $E_{2g}^2$ (P3), and $A_{1g}^1$ (P4), scale bars: 5 μm. **(c)** Temperature-dependent spectra of CuS crystal (80-300 K, step: 20 K). **(d)** Raman peak positions of CuS (P1-4) as a function of the measured temperature.

High-resolution transmission electron microscopy (HR-TEM), selected-area electron diffraction (SAED), and energy-dispersive X-ray spectroscopy (EDX) were all employed to better comprehend the CuS crystal's structure and composition. **Figure 3a** displays the results of an HR-TEM that utilized Fast Fourier Transform filtering to get a precise image of the atomic arrangement in the CuS crystal. The image showed a highly organized structure, with atoms arranged in a regular hexagonal lattice. Notably, between two intersecting planes at 120°, representing crystallographic planes (100) and (010) of the hexagonal crystal structure, the interplanar gap measured 0.35 nm. This verified that the CuS material indeed has the predicted hexagonal crystal structure. The SAED structural analysis (**Figure 3b**) revealed a 6-fold symmetric diffraction pattern along the [001] zone axis, consistent with the HR-TEM results. Consistent with the pole figure, the SAED pattern shows evidence of a superlattice or stacking fault in the CuS crystal, with a secondary 6-fold symmetry (lesser intensity with some of the spots highlighted in a red circle). Such phenomena have been widely reported to exist in various low-dimensional materials.[37-40] Complementary to the structural analyses, EDX elemental analysis was used to determine the CuS crystal's atomic composition (**Figure 3c**). The EDX analysis showed that the sample contained both Cu and S. Elemental mapping (**Figure 3d**) also showed that Cu and S atoms were evenly distributed across the crystal, confirming the stoichiometric distribution of the elements in CuS crystals. Moreover, X-ray photoelectron spectroscopy (XPS) was performed to determine the chemical composition and the surface electronic states of as-obtained crystals. As shown in **Figure 3e**, the Cu 2p peaks are deconvoluted into two separate overlapping doublet peaks using the Gaussian fitting curves. The doublets are attributed to Cu $2p_{3/2}$ and Cu $2p_{1/2}$ of the $Cu^{2+}$ oxidation states. The peaks at binding energies of 931.9 eV (Cu $2p_{3/2}$) and 951.6 eV (Cu $2p_{1/2}$) belong to the $Cu^{2+}$ oxidation state in the pure CuS crystal. While the remaining two peaks at binding energies of 933.4 eV and 953.6 eV correspond to Cu $2p_{3/2}$ and Cu $2p_{1/2}$ in $Cu_{1-x}S$, indicating that defects may exist in the crystals.[41, 42] Besides, the peaks centered at binding energies of 164.5 eV and 166.5 eV are ascribed to S $2p_{3/2}$ and S $2p_{1/2}$, respectively (**Figure 3f**).

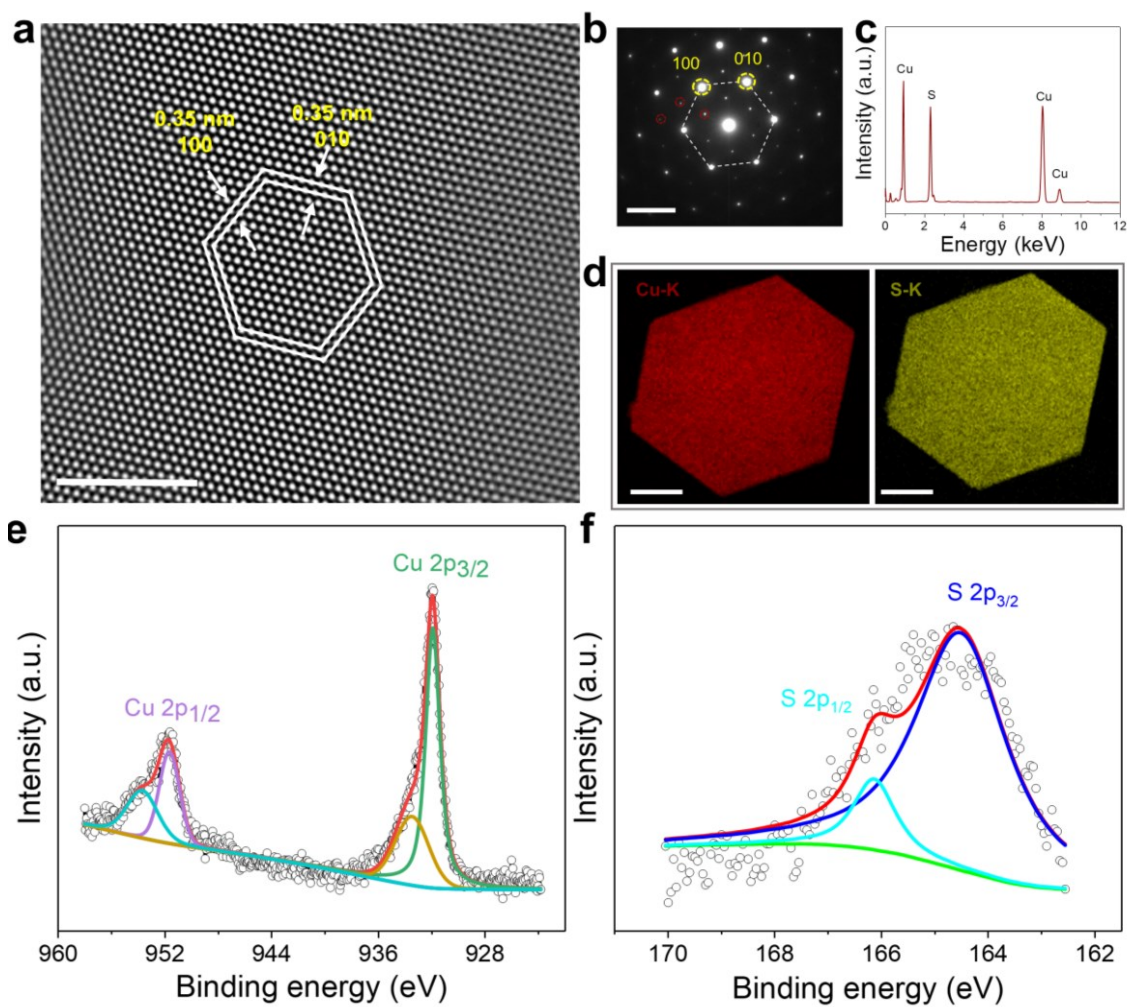

**Figure 3:** Structural and chemical compositional characterization of the CuS crystal. **(a)** High-resolution TEM image. **(b)** The SAED patterns. **(c)** EDX spectrum. **(d)** Elemental mapping for Cu and S, respectively. XPS spectra of **(e)** Cu 2p, and **(f)** S 2p.

Next, we focus on the Second Harmonic Generation (SHG) of the CuS. SHG is an effective approach to investigate physical phenomena associated with crystal symmetry, and has an exceptional sensitivity to materials' electronic and structural properties.[43-46] The centrosymmetric crystal structure of CuS may be disrupted by various factors such as defects or stalking faults, which could lead to the formation of multi-oriented domains within the crystal, and ultimately trigger a SHG response.[47, 48] The results of our EBSD, SEAD, and XPS analyses indicate the presence of structural defects within the CuS lattice. These defects may be responsible for disrupting the crystal's centrosymmetry. Thus, SHG was utilized to corroborate it further. The SHG mechanism is depicted in **Figure 4a**, where the incident laser (ω) generates an (2ω) response. The SHG response of an as-synthesized CuS crystal under various incident laser wavelengths from the edge of visible light to near-infrared (760 to 1020 nm) is presented in **Figure 4b**, which shows a wide spectrum response with distinct wavelength selectivity. Moreover, the SHG mappings display a uniform response throughout the entire CuS lattice (**Figure 4b** inset). Evolution of SHG intensity with changing incident laser power was also further systematically investigated. With increasing the incident laser power from 0.7 to 1.6 mW under 800 nm laser excitation, the intensity of the SHG signal at 400 nm exhibits significant enhancement (**Figure 4c**). The relationship between SHG intensity and laser power was fitted linearly in the log-log coordinate, as displayed in **Figure 4d**. Interestingly, the slope of 2.05 is close to the theoretical value of 2 calculated from the electric dipole theory.[49]

Before assessing polarization, we rotated the sample to a position where the highest SHG response could be generated by setting the initial azimuthal angle to 0°. In parallel (XX) and perpendicular (XY) directions, the typical 6-fold symmetry pattern fitted proportionally with $\sin^2 3\theta$ and $\cos^2 3\theta$ can be detected, as presented in **Figure 4e-f**. A similar set of SHG studies was conducted on a different CuS crystal exhibiting different morphology, which demonstrated a commensurate SHG response, as illustrated in Figure S3. It implied broken inversion symmetry that is characteristic of hexagonal-symmetric structures similar to other SHG sensitive materials. Due to this unique quality, CuS crystal possesses several intriguing features with considerable application potential in the field of nonlinear optics. Thus, CuS crystal's promise in nonlinear optics increases the variety of materials available for these applications and presents new avenues for investigation into improving its performance and uncovering unanticipated capabilities.

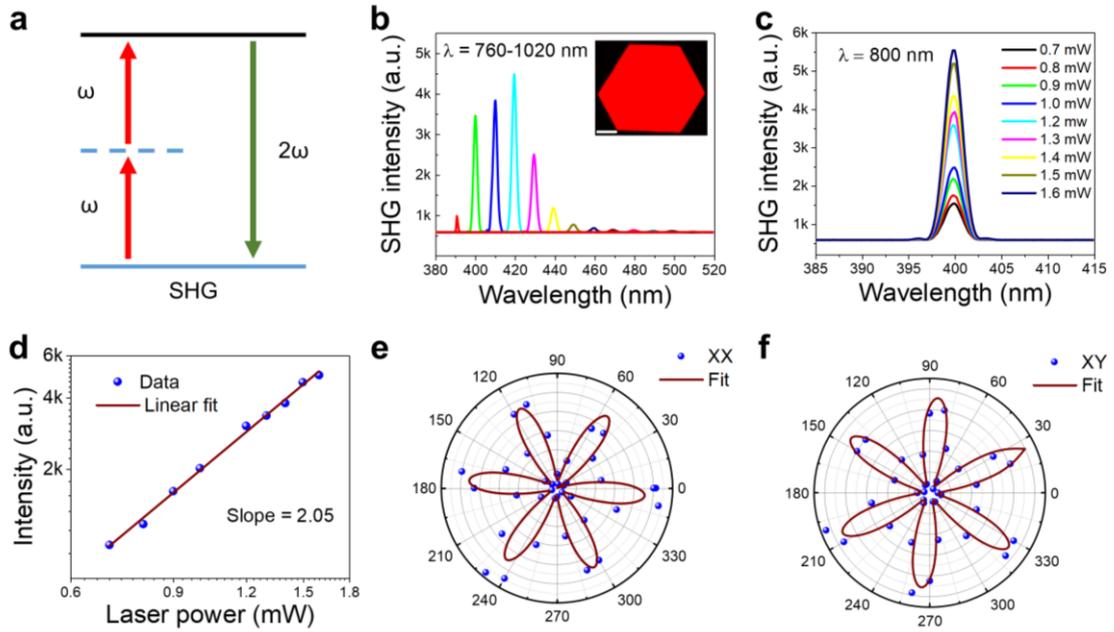

**Figure 4:** SHG characterization of CuS crystal. **(a)** Basic mechanism of SGH. **(b)** The SHG spectra of CuS crystal under various excitation wavelengths (760 - 1020 nm). Inset is the SHG mapping of CuS crystal under 800 nm laser excitation, scale bar corresponds to 3 μm, **(c)** The SHG spectra of the CuS crystal with different incident powers. **(d)** The SHG intensities as a function of incident power. **(e-f)** Polarization angle-dependent SHG intensity under parallel and perpendicular polarization configurations, respectively.

**Conclusion**

In conclusion, CuS crystals were successfully synthesized for the first time using a single-step CVD technique. The as-grown CuS crystals were characterized and crystals as thin as14 nm up to a lateral size of up to 60 μm are demonstrated. Temperature dependence of Raman shifts are measured. An unexpected SHG response was observed in the as-synthesized CuS crystal owing to the presence of intrinsic defects in the lattice structure. These defects caused a breakdown in local symmetry in the CuS lattice, thereby revealing the material's SHG behavior. Our findings suggest that CuS could find new uses in nonlinear optical applications beyond its current utilization in catalysis and electronics.

**Materials and Characterization**

**CVD growth:** CuS crystals were grown in a tubular furnace with a single temperature zone and atmospheric pressure CVD conditions. A quartz boat containing a CuCl powder (97%, Sigma Aldrich) was placed in the middle of the temperature zone. S powder (99.5%, Sigma Aldrich) was inserted at the upstream end of the tube, and the temperature was maintained at 200. Substrates, e.g., cleaved fluorphlogopite mica, were positioned 8 cm apart from the furnace's center in the downstream position. The tube was pumped and cleaned with 500 sccm Ar flow to drain air prior to heating. Then, the furnace was heated to 600 °C at a rate of 30 ºC/min using steady 50 sccm Ar as the carrier gas, and it was held at that temperature for 30 minutes. After the procedure was concluded, the furnace was allowed to cool naturally.

**Characterizations:** CuS crystal morphologies were examined using an OM (BX51, OLYMPUS) and an AFM (Bruker Dimension Icon). The crystalline structure, orientation, and composition were investigated using XRD ($\lambda$: 1.54 Å, D2 phaser, Bruker), XPS (AXIS-ULTRA DLD-600W, Kratos), EBSD (FEI Quanta650), and TEM (Tecnai G30 F30, FEI). Raman spectra were acquired using a confocal Raman system (Alpha 300R, WITec) equipped with a 532 nm laser.

**SHG measurements:** SHG measurements were performed in an (alpha300RS+, WITec) Raman system with a reflection mode under normal incidence excitation using a femtosecond laser as the excitation source. A mode-locked Ti: sapphire laser with a pulse duration of 140 fs and repetition rate of 80 MHZ generated the output laser with a continually varying wavelength ranging from 340 nm to 1600 nm, which was then filtered into an optical parametric oscillator (Chameleon Compact OPO-Vis). A dichroic beam splitter was used to reflect the laser beam into the 100x objective lens with a spot size of roughly 1.8 µm and communicate the reflected SHG signal. The reflected SHG signal was then filtered with a short pass (SP) filter before being sent to the spectrometer and CCD. The collected polarized SHG signal was sent through a linear polarized analyzer for SHG polarization measurement by rotating the sample with a step of 10° relative to fixed light polarization. All experiments were carried out in a natural setting.

## ASSOCIATED CONTENT

## AUTHOR CONTRIBUTION

A.A.S: Synthesis, characterizations, conceptualization, data curation, writing-original draft, co-corresponding. R.R: Editing. A.P: Characterizations, editing. T.S.K: Supervision, conceptualization, funding, editing, corresponding. All authors have agreed on the final version of the manuscript.

## CONFLICT OF INTEREST

No competing financial interests are declared.

## ACKNOWLEDGMENTS

The authors acknowledge funding from the Scientific and Technological Research Council of Turkey (TUBITAK) under grant number 120N885.

SUPPORTING INFORMATION

# Second Harmonic Generation in Chemical Vapor Deposition Synthesized CuS Crystals

Abdulsalam Aji Suleiman[a]*, Reza Rahighi[a], Amir Parsi[a], and Talip Serkan Kasirga[a,b]*

[a]Institute of Materials Science and Nanotechnology, Bilkent University UNAM, Ankara 06800, Turkey

[b]Department of Physics, Bilkent University, Ankara 06800, Turkey

*Corresponding authors;

Email: kasirga@unam.bilkent.edu.tr; abdulsalam@unam.bilkent.edu.tr

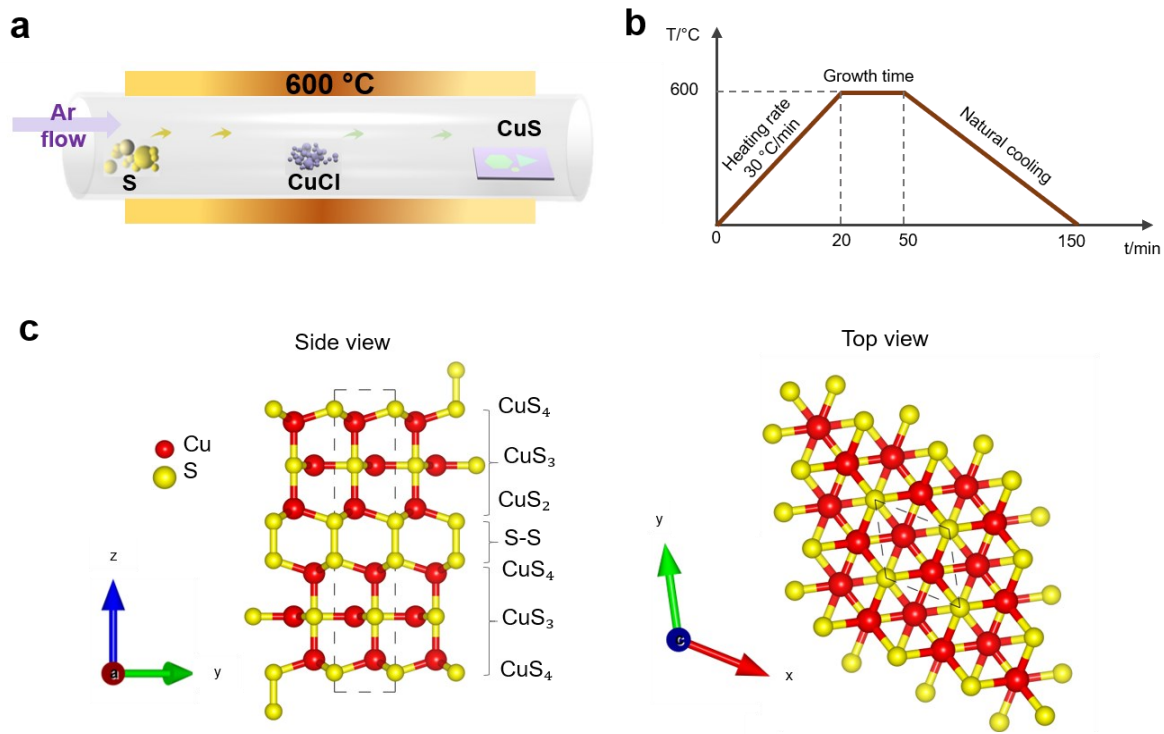

**Figure S1: (a)** The schematic image of the CVD setup. **(b)** The growth profile of the CVD-grown crystals. **(c)** The schematic side-view (left) and top-view (right) of the CuS crystal structure.

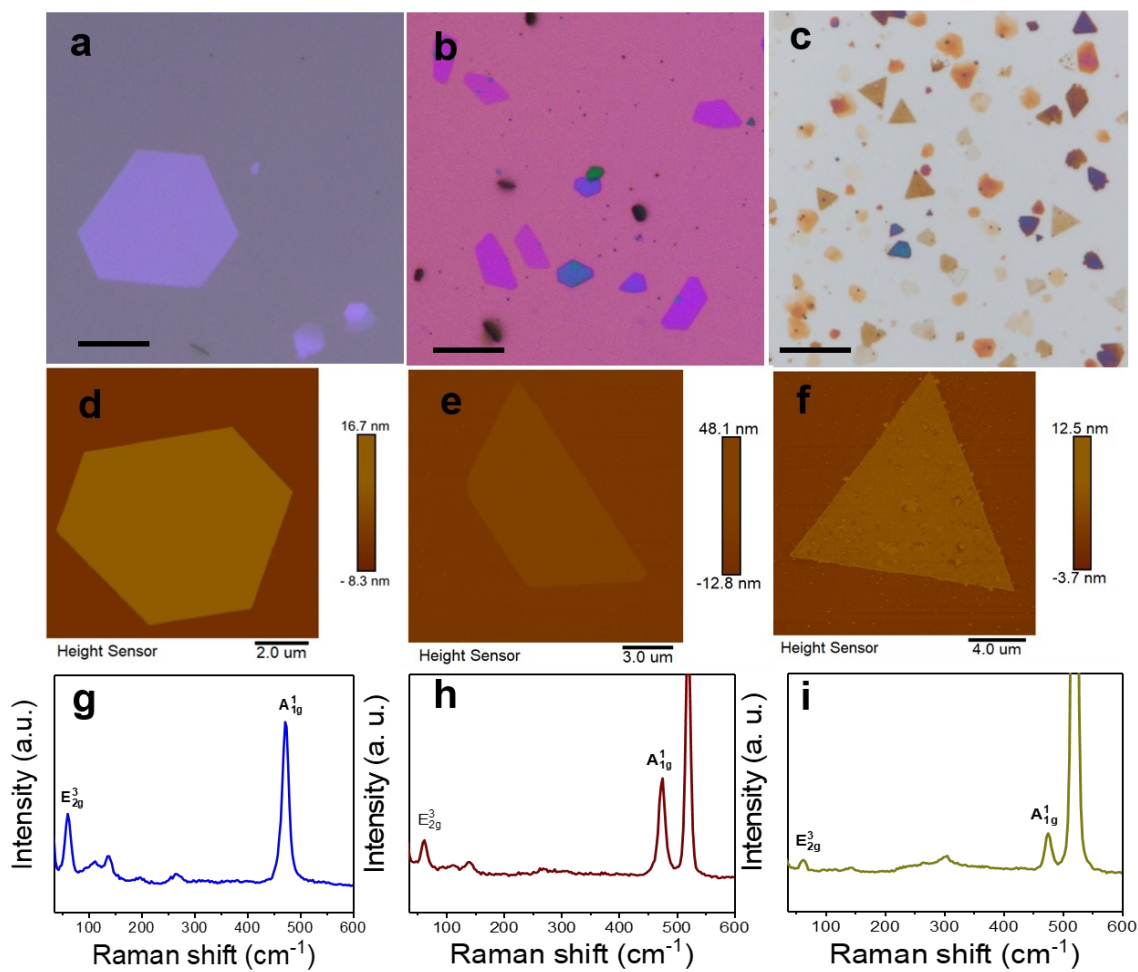

**Figure S2: (a-c)** Optical microscope images of the CuS crystals grown on various substrates of mica, Si/SiO$_2$, and Si, respectively. The scale bars are 10, 20, and 20 μm, respectively. **(d-f)** Corresponding AFM images of the CuS crystals with their height profiles of 40, 28, and 6 nm, respectively. **(g-i)** Raman spectra of the CuS crystals grown on mica, Si/SiO$_2$, and Si, respectively.

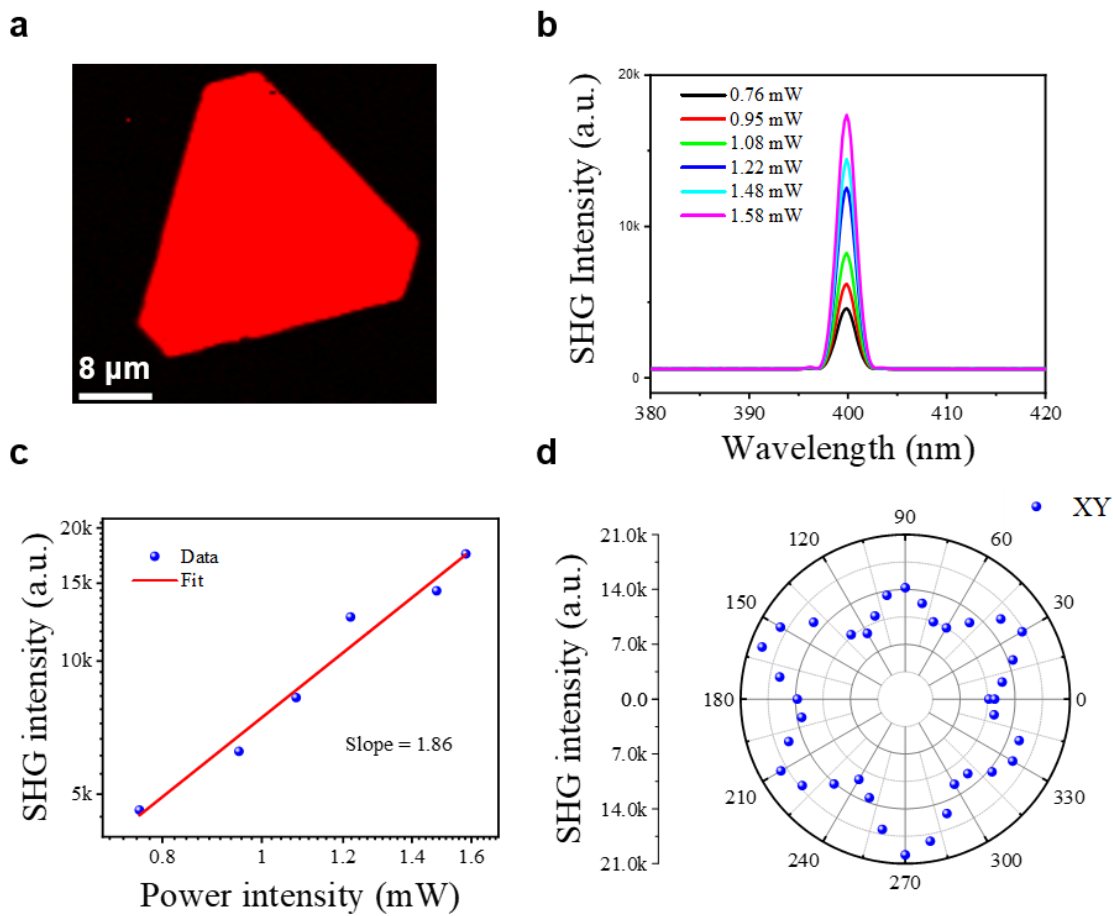

**Figure S3: (a)** The SHG mapping of the CuS crystal under 800 nm laser excitation. **(b)** The SHG spectra of the CuS crystal with different incident powers. **(c)** Incident power-dependent SHG intensities. **(d)** Polarization angle-dependent SHG intensity under perpendicular polarization configurations (The excitation laser is 800 nm with a power of 1.2 mW).